\documentclass[aps,twocolumn,groupedaddress,numberedheadings]{revtex4}

\usepackage{times}
\usepackage{mathrsfs}           
\usepackage{textcomp}           
\usepackage{graphicx}           
\usepackage{braket}             
\usepackage{amsmath,amsthm,amsfonts,amscd,amssymb}
\usepackage{url}                

\newcommand{\carb}{CO$_2$}      
\newcommand{\et}{{\em et~al.}}  
\newcommand{\cm}{cm$^{-1}$}     
\newcommand{\sgp}{$\displaystyle\Sigma_{\mbox{\scriptsize g}}^{\mbox{\tiny
        +}}$}
\newcommand{\dg}{$\displaystyle\Delta_{\mbox{\scriptsize g}}$}

\begin{document}

\title{Raman observations of quantum interference in the \\
        $\nu_1 / 2\nu_2$ Fermi dyad region of carbon dioxide}

\author{Craig W. McCluskey}
\altaffiliation{Currently with: Los Alamos National Laboratory,
				N-2, Advanced Nuclear Technology,
				P.O. Box 1663, M/S B228,
				Los Alamos, NM 87545 USA}
\email[]{craigm@lanl.gov}
\affiliation{Department of Physics C1600, The University of Texas at Austin,
		Austin, TX  78712-0264, USA}

\author{David S. Stoker}
\affiliation{Department of Physics C1600, The University of Texas at Austin,
		Austin, TX  78712-0264, USA}
\affiliation{ }
\affiliation{The Center for Nano- and Molecular Science and Technology
		and the Texas Materials Institute, C2201 \\
The University of Texas at Austin, Austin, TX 78712-1063, USA}


\date{\today}

\begin{abstract}

Coherent anti-Stokes Raman spectra (CARS) were obtained for CO$_2$
    in a positive column discharge.
The intensities of the Raman transitions to the $\nu_1$/$2\nu_2$ Fermi
    dyad decrease significantly when the discharge is turned on because
    of equalization of the lower and upper state populations by electron
    impact excitation, in a manner similar to saturation.
The strength of the 1285~cm$^{-1}$ transition is observed to decrease
    by a factor of 20 greater than the 1388~cm$^{-1}$ line due to a
    quantum interference decreasing the vibrational relaxation rate
    of the upper state of the 1285~cm$^{-1}$ transition.
This interference is verified by measurements of the decay rates from
    the state.
Experiments ruling out Stark effects and polarization effects are described.
Supporting the observed rapid vibrational excitation by electrons,
    a strong hot band transition $20^00$ $\leftarrow$ $02^20$ at
    1425.61~cm$^{-1}$ was observed.
These observations are compared with recent measurements of the Raman
    spectra of CO$_2$ heated in a flame.

\end{abstract}

\maketitle

\section{Introduction \label{sec:intro.CO2}}

\carb\ has been studied since the earliest days of both infrared
  and Raman spectroscopy.
Rasetti wrote about the Raman effect in gases \cite{Rasetti:gases}
  and selection rules in the Raman effect \cite{Rasetti:selection}
  in 1929. He had noticed, in particular, that the Raman lines
  observed in \carb\ were quite different than expected, with two
  lines instead of a single line.
Fermi \cite{Fermi:dyad} described the origin of these strong Q-branch
  lines which now bear his name in 1931.
Discussions about quantum interference in transitions between the
  states of \carb\ waited until much later. Zhu \et\ \cite{Zhu:quant.int},
  for example, wrote about quantum interference in the excitation rates
  in 1991.
Quantum interference effects are also observable in relaxation and
  are the subject of this work.


\section{Theory \label{sec:theory.CO2}}


\subsection{Anharmonicity in \carb\ \label{subsec:CO2anharm}}

The Q-branches in \carb\ at 1285~\cm\ and 1388~\cm\ are the
components of the Fermi dyad, a feature due to an ``accidental''
near degeneracy of one quanta of energy in symmetric stretch,
$\nu_1$, with two quanta of energy in bend, $2\nu_2$ \cite{Fermi:dyad}.
Because these two frequencies are so close to each other
($\delta = 7.87$~\cm\ \cite{Howard-Lock:rimfd}) the states
$\Ket{10^00}$ and $\Ket{02^00}$ strongly perturb each other.

As Herzberg \cite{Herzberg:vol.2} shows, anharmonicities change the
total energy of a polyatomic molecule from the sum of $3N$ mutually
independent simple harmonic oscillators of mass 1, to,
\begin{eqnarray}
\mathscr{H} & = &
  \frac{1}{2}\,(\dot{\eta}_{1}^{\,2} + \lambda_{1}\eta_{1}^{\,2}) +
  \frac{1}{2}\,(\dot{\eta}_{2}^{\,2} + \lambda_{2}\eta_{2}^{\,2}) +
  \cdots \nonumber				\\
	    &   &
  + \sum_{ijk}\alpha_{ijk}\eta_{i}\eta_{j}\eta_{k} +
  \sum_{ijkl}\beta_{ijkl}\eta_{i}\eta_{j}\eta_{k}\eta_{l} + \cdots  .
\label{eqn:anharm}
\end{eqnarray}
where $\eta_i$ are the normal coordinates of the motions. Thus,
the total energy is no longer the sum of independent (even though
anharmonic) oscillators.

In \carb\ some of the coefficients $\alpha_{ijk}$ and
$\beta_{ijkl}$ are equal to zero because the potential energy must
be unchanged for all symmetry operations of the molecule's point
group, \textbf{D}$_{\infty h}$. Additionally, symmetry operations
allow the antisymmetric normal coordinates to occur in
Eqn.~\ref{eqn:anharm} only in even powers. Of the ten unique cubic
terms, then, only $\alpha_{111}$, $\alpha_{122}$, and
$\alpha_{133}$ are non-zero and the cubic terms of the potential
energy are,
\begin{equation}
\alpha_{111}^{ }\eta_{1}^{\,3} +
\alpha_{122}^{ }\eta_{1} (\eta_{2a}^{\,2} + \eta_{2b}^{\,2}) +
\alpha_{133}^{ }\eta_{1}^{ }\eta_{3}^{\,2} \; .
\label{eqn:cubic}
\end{equation}
$\alpha_{122}$ is the term that generates the anharmonic coupling
between $\nu_1$ and $2\nu_2$ that gives the Fermi resonance.

Note that in Raman spectroscopy, the signal is proportional, among
other things, to the derivative of the polarizability with respect
to position at the equilibrium distance. Because of this, the
symmetric stretch normal mode ($\nu_1$) produces a Raman signal,
while bend ($\nu_2$) and assymetric stretch ($\nu_3$) are Raman
inactive.


\subsection{Fermi Splitting in \carb\ \label{subsec:CO2fermisplit}}

As Herzberg \cite{Herzberg:vol.2} also shows, in Fermi resonance
the two vibrational levels that have nearly the same energy in
zero approximation (in \carb\ $\nu_1$ and $2\nu_2$) ``repel'' each
other: one is shifted up and the other is shifted down so that the
separation of the two levels is much greater than expected. In
addition, a mixing of the eigenfunctions of the two states occurs.
The deviation of the energy levels from what is expected, as well
as the mixing of the eigenfunctions, become greater as the
separation of the zero approximation energies becomes smaller.

In addition to depending inversely on the separation of the zero
approximation energies, the magnitude of the repulsion depends on
the value of the corresponding matrix element $W_{ni}$ of the
perturbation function $W$:
\begin{equation}
W_{ni} = \int \psi_{n}^{\,0} \, W  \psi_{i}^{\,0*}  d\tau
\label{eqn:wni}
\end{equation}
where W is essentially given by the anharmonic (cubic, quartic,
etc.) terms of the potential energy discussed in
\S\ref{subsec:CO2anharm} and $\psi_{n}^{\,0}$ and $\psi_{i}^{\,0}$
are the zero approximation eigenfunctions of the two levels. Note
that since W has the full symmetry of the molecule, as mentioned
above, $\psi_{n}^{\,0}$ and $\psi_{i}^{\,0}$ must both have the
same symmetry type in order for the integral in (\ref{eqn:wni}) to
be non-zero. This means that only vibrational levels of the same
symmetry can perturb each other.  This can be seen in the energy
level diagram, Fig.~\ref{fig:CO2el}, where $10^00$ and $02^00$,
both being \sgp, are involved in a Fermi resonance while $02^20$,
being \dg, is not.

If the separation of the zero approximation energies is fairly
small, the magnitude of the shift can be calculated with
first-order perturbation theory from the secular determinant,
\[
\left |
\begin{array}{cc}
	E_n^{\,0} - E	&	W_{ni}		\\[8pt]
	W_{in}		&	E_i^{\,0} - E
\end{array}
\right |	\; = \; 0	\; ,
\]
where, $E_n^{\,0}$ and $E_i^{\,0}$ are the unperturbed energies
and $W_{in}^{ } = W_{ni}^{\,*}$ from Eqn.~(\ref{eqn:wni}). The
perturbed energies are then,
\begin{equation}
E = \, \overline{\!E}_{ni} \pm \textstyle\frac{1}{2}
		\displaystyle\sqrt{4\,|W_{ni}|^2 + \delta^2\,} \; ,
\end{equation}
where,
\[
\begin{array}{rcl}
\overline{\!E}_{ni} & = & \frac{1}{2}\,(E_n^{\,0} + E_i^{\,0})
					\\ [7pt]
\delta & = & (E_n^{\,0} - E_i^{\,0})			\quad .
\end{array}
\]
The eigenfunctions of the two resulting states are linear
combinations of the zero approximation eigenfunctions of the form,
\begin{eqnarray}
\psi_n & = & a \psi_n^{\,0} - b \psi_i^{\,0}	\nonumber	\\
\psi_i & = & b \psi_n^{\,0} + a \psi_i^{\,0} \; ,
\label{eqn:lincombo}
\end{eqnarray}
where,
\begin{eqnarray}
a & = & \left (
	\frac
	{\sqrt{4\,|W_{ni}|^2 + \delta^2 \,} + \delta \,}
	{2 \sqrt{4\,|W_{ni}|^2 + \delta^2\,} }
	\right )^\frac{1}{2}
, \nonumber \\
b & = & \left (
	\frac
	{\sqrt{4\,|W_{ni}|^2 + \delta^2 \,} - \delta \,}
	{2 \sqrt{4\,|W_{ni}|^2 + \delta^2\,} }
	\right )^\frac{1}{2}
\; .
\label{eqn:coeff}
\end{eqnarray}

Putting Howard-Lock and Stoicheff's values \cite{Howard-Lock:rimfd},
\begin{eqnarray*}
W_{10^00 - 02^00} & = & - \,  51.232~\textrm{~cm}^{-1} \nonumber	\\
\delta & = &  - \, 7.87~\textrm{~cm}^{-1}	\; ,
\end{eqnarray*}
in Equations \ref{eqn:coeff} (using the magnitude of $\delta$) gives,
\[
a = 0.734 \; , \qquad b = 0.679 \; .
\]
Identifying $\psi_n^{\,0}$ with $\Ket{10^00}$ and $\psi_i^{\,0}$
with $\Ket{02^00}$, expressing the coefficients as sines and cosines,
and using bracket notation, as do Zhu~\et \cite{Zhu:quant.int},
yields,
\begin{eqnarray}
\Ket{\psi_l} & = & \cos{\theta}\Ket{10^00}
			- \sin{\theta}\Ket{02^00}	\nonumber \\
\Ket{\psi_u} & = & \sin{\theta}\Ket{10^00}
			+ \cos{\theta}\Ket{02^00}	\; ,
\label{eqn:psiul}
\end{eqnarray}
with $\theta~=$~42.8\textdegree\ for $^{12}$\carb\

The magnitude of $\delta$ must be used in Equations~\ref{eqn:coeff}
because if the minus sign from $\delta = -\,7.87$~\cm\ were used, the
lower state, $\Ket{\psi_l}$, would have an excess $\Ket{02^00}$ instead
of $\Ket{10^00}$, which is the unperturbed state which is actually
``repelled'' toward it.


\subsection{Notation Conventions \label{subsec:CO2notation}}

One of the problems in working with \carb\ is the confusion
engendered by the profusion of state naming and notation
conventions.
Adel and Dennison \cite{Adel:CO2} used the notation
  (v$_1$,~v$_2$,~v$_3$,~$\ell$), where $\ell$ is the vibrational
  angular momentum.
Herzberg \cite{Herzberg:vol.2} used the notation
  (v$_1$,~v$_2\/^{\ell}$,~v$_3$).
Amat and Pimbert \cite{Amat:CO2} used notation similar to Herzberg's
  but introduced the idea of using 1 and 2 as ranking indices to
  denote the higher and lower-energy peaks of a Fermi dyad.
Schr{\"o}tter and Brandm{\"u}ller in
  Finsterh{\"o}lzl~\et~\cite{Schroetter:carb.raman.I} used the
  notation of Amat and Pimbert, but explicitly labeled states
  with the ranking index.  They, for example, labeled the
  1388~\cm\ peak as $(10^{0}0)_{\textrm{I}}$ and the
  1285~\cm\ peak as $(10^{0}0)_{\textrm{II}}$. They
  also labeled the 1410 hotband as
	$(11^{1}0)_{\textrm{I}}-(01^{1}0)$
  and the 1265 hotband as
	$(11^{1}0)_{\textrm{II}}-(01^{1}0)$,
  keeping for the hotbands some indication of which levels
  are involved.
Rothman and Young used notation similar to that of Schr{\"o}tter and
  Brandm{\"u}ller in their HITRAN (\textbf{hi}gh resolution
  \textbf{trans}mission) molecular absorption database intended
  for atmospheric physicists, but wrote everything out on a single
  line suitable for a computerized database \cite{HITRAN:url}.
  In their notation, the 1388~\cm\ line is 10001 and the 1285~\cm\
  line is 10002.

\begin{figure*}[t]  
\includegraphics[width=5in,height=!]{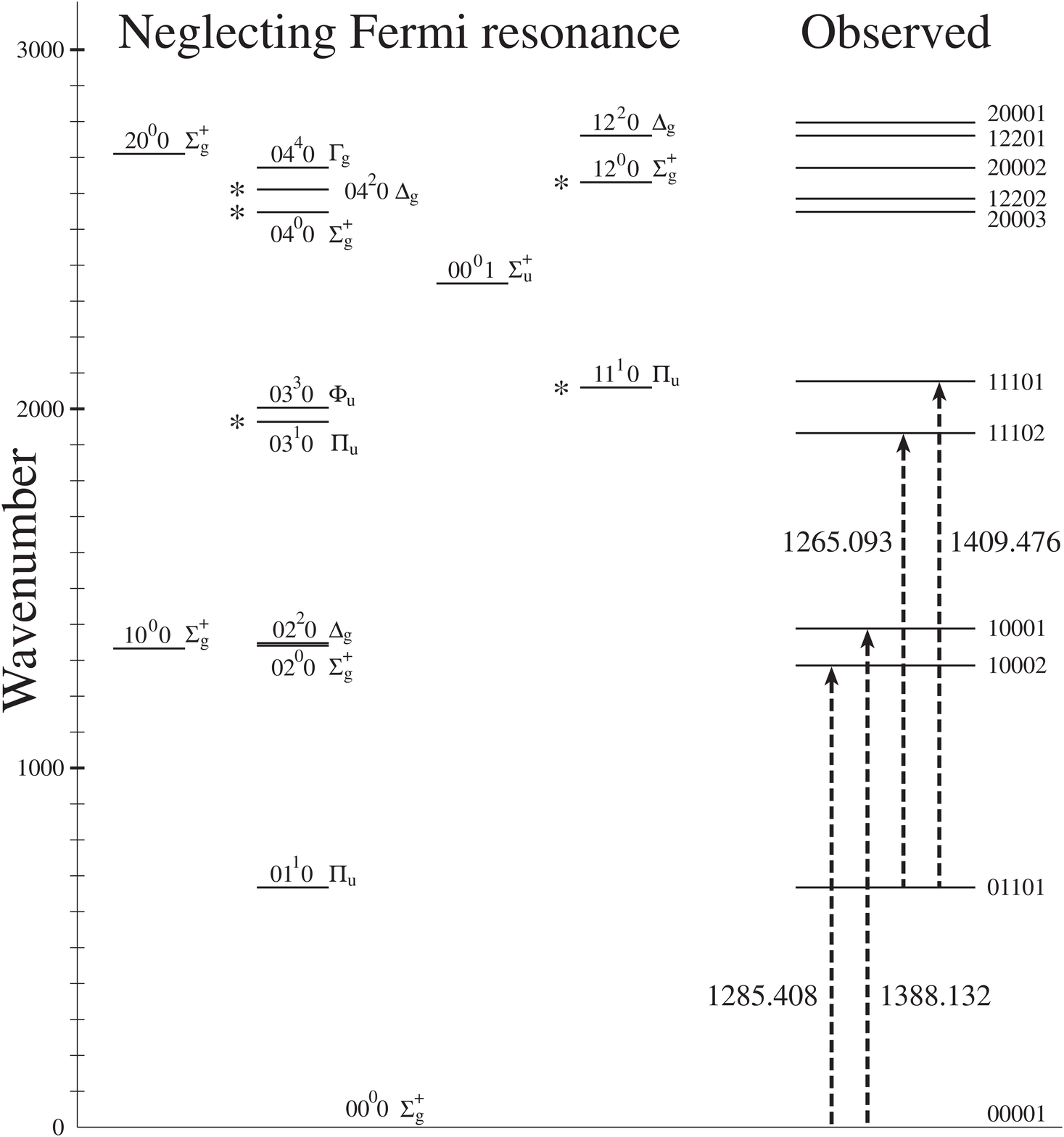}
\caption{Energy Levels of \carb. Energies of levels marked with an
	asterisk are uncertain. Strongest Raman transitions are marked
	by heavy dotted lines. \label{fig:CO2el}}
\end{figure*}
All this confusion is {\em somewhat} ameliorated by the
``Rosetta Stone'' of notation in Rothman and Young's 1981
paper on carbon dioxide \cite{Rothman:carb.ir.levels.II},
reproduced below in Table~\ref{table:rosetta}.
\begingroup
\begin{table}[hbt]
\caption{Comparison of notations for \carb\ vibrational energy levels.
		\label{table:rosetta}}
\begin{ruledtabular}
\begin{tabular}{ccc}
Herzberg \cite{Herzberg:vol.2} & Amat \cite{Amat:CO2} & HITRAN \cite{AFGL} \\
\cline{1-3}
$20^{0}0$ & $(20^{0}0,04^{0}0)_{\textrm{I}}$   & 20001 \\
$12^{0}0$ & $(20^{0}0,04^{0}0)_{\textrm{II}}$  & 20002 \\
$04^{0}0$ & $(20^{0}0,04^{0}0)_{\textrm{III}}$ & 20003 \\
$12^{2}0$ & $(12^{2}0,04^{2}0)_{\textrm{I}}$   & 12201 \\
$04^{2}0$ & $(12^{2}0,04^{2}0)_{\textrm{II}}$  & 12202 \\
$04^{4}0$ & $04^{4}0$                          & 04401 \\
\end{tabular}
\end{ruledtabular}
\end{table}
\endgroup

As one progresses from Herzberg's notation to the HITRAN notation,
it becomes increasingly difficult to know which exact levels are
involved and the states involved in the transition become hidden.

The vibrational energy levels of \carb\ are shown in Fig.~\ref{fig:CO2el}.
In this figure, the strongest Raman transitions of \carb\ in the
wavenumber range examined are marked by heavy dotted lines.
Herzberg's notation is used on the left to show the locations of
where the states would be if they were not perturbed by Fermi
resonance. Some of the energy levels of these lines, marked by
asterisks, were given in Herzberg \cite{Herzberg:vol.2} but were
not given in any of the other references examined.  Herzberg's
(old) values of the energies did not make any sense considering
Fermi resonance and the (new and more precise) observed levels,
which are plotted on the right according to Rothman's values
\cite{Rothman:energylevels} and marked with HITRAN notation, so
the values of these levels were manually adjusted to be more
reasonable.  Note that in Fig.~\ref{fig:CO2el}, the non-Fermi
resonance influenced levels of $02^20$, $03^30$, and $04^40$
are not plotted on the right side for clarity.


\section{Experimental Apparatus and Conditions \label{sec:exp.CO2}}

\begin{figure}[htb]  
\includegraphics[width=3.3in,height=!]{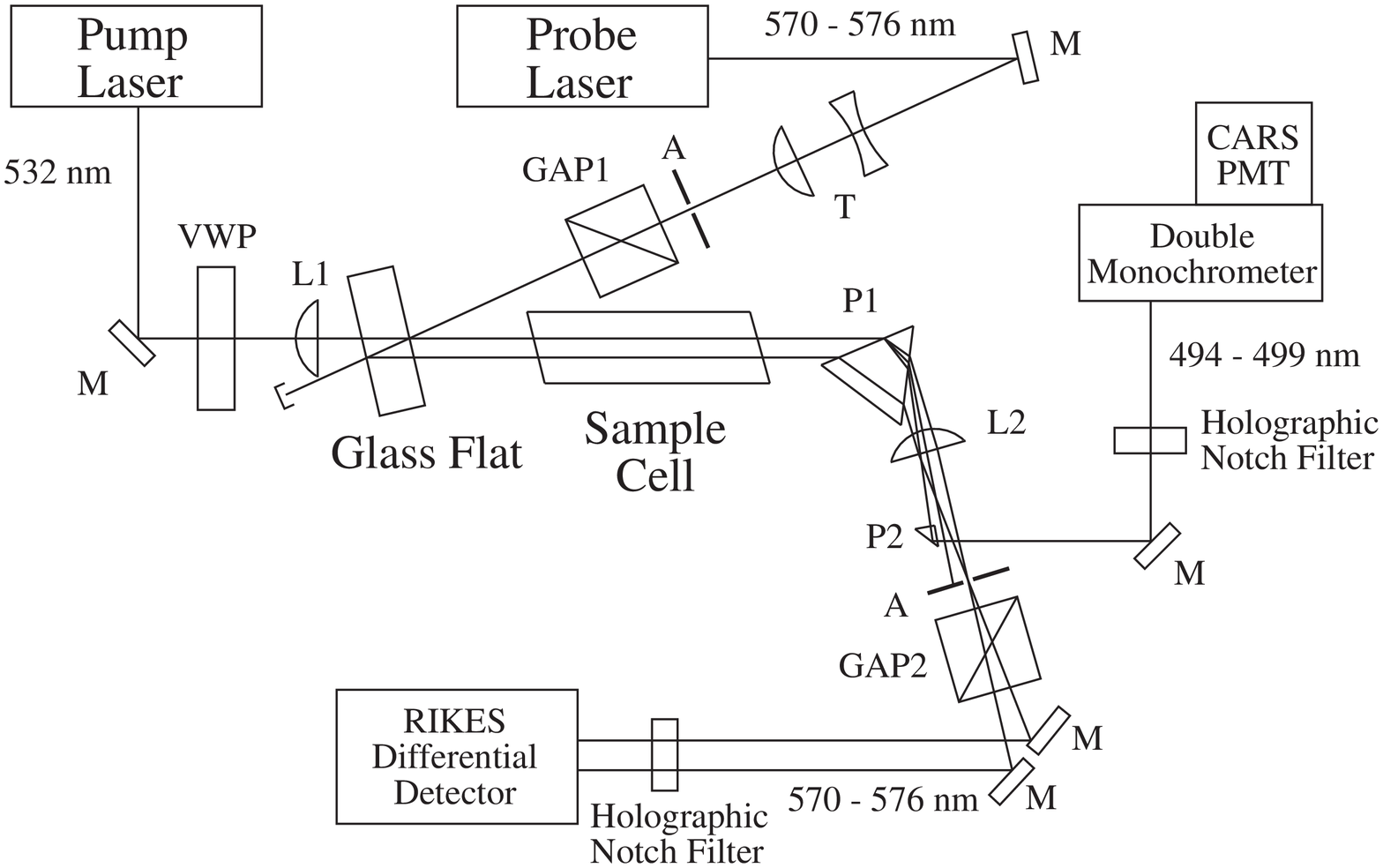}
\caption{Spectrometer Optics Layout. A: aperture, GAP: Glan-air polarizer,
	L: lens, M: mirror, P: prism, T: Galilean telescope, VWP: Variable
	wave plate.
\label{figure:opticslayout}}
\end{figure}

The experimental setup used for this work, described previously in the
literature \cite{Keto:rikes}, is shown in Fig.~\ref{figure:opticslayout}.
The spectrometer consists of a neodymium-YAG laser (using the second
harmonic output at $\lambda=532$~nm), a tunable dye laser pumped by part
of the output of the YAG laser, optics to overlap both spatially and
temporally the remainder of the YAG laser's output (which is represented
by the box labeled ``Pump Laser'' in Fig.~\ref{figure:opticslayout}) with
the output of the dye laser inside the sample volume, and optics to detect
the newly created CARS beam coming from the sample cell. A quartz block in
a squeezing mechanism is used as a variable wave-plate (VWP) to adjust the
polarization of the incident YAG beam.  Fig.~\ref{figure:opticslayout} also
shows optics and electronics for detection of Raman induced Kerr-effect
signals, which were used previously \cite{Keto:rikes}. For this experiment,
the ``pick-off'' prism (P2 in Fig.~\ref{figure:opticslayout}), the double
monochromator, and the photomultiplier tube (PMT) were added to detect the
coherent anti-Stokes Raman spectroscopy (CARS) signal.

\clearpage

\begin{figure}[hbt]  
\includegraphics[width=3.3in,height=!]{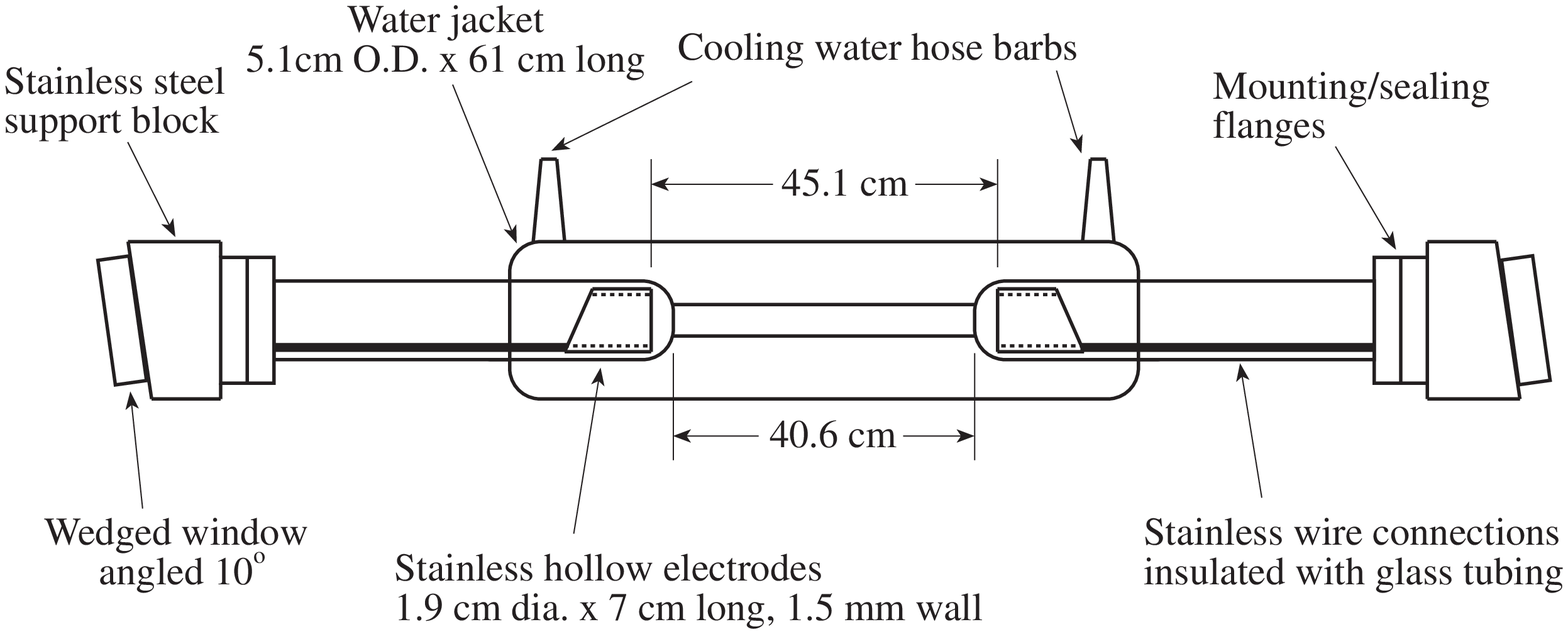}
\caption{The Linear Discharge Cell. \label{figure:LDC1}}
\end{figure}

The Pyrex glass linear discharge cell used for the majority of the
experiments is shown in Fig.~\ref{figure:LDC1}. It consists of two
concentric glass tubes and has both a hollow cathode and a hollow
anode.
The inner glass tube is 1$''$~O.D. at the ends and necks down
to an inner diameter of 0.375$''$ over its central 16~inches.
The smaller cross-section confines the glow discharge and
yields a higher current density. The ends of the inner glass
tube are open for mounting to stainless steel support blocks
at each end.  The outer glass tube forms a water-cooling jacket
which cools not only the region of the discharge, but also
beyond the electrodes.

When running, the cell provides a long positive column glow
discharge which extends four to five~inches on both sides
of the focus of the YAG and dye laser beams so the region
probed by the most intense portions of the focused laser
beams is within the positive column.  At a current of 100~mA,
the current density is 1400~A/m$^2$ (0.14~A/cm$^2$).


\section{Results \label{sec:results.CO2}}

\begin{figure}[htb]  
\includegraphics[width=3.3in,height=!]{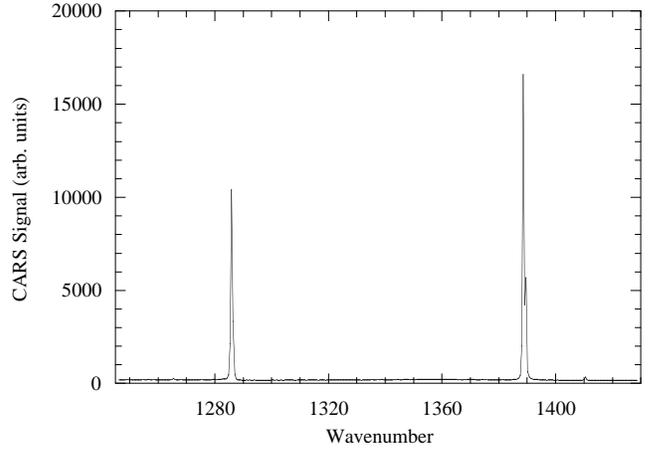}
\caption{Spectrum of carbon dioxide with discharge off.
		\label{fig:CO2.ldcnd}}
\end{figure}

\begin{figure}[hbt]  
\includegraphics[width=3.3in,height=!]{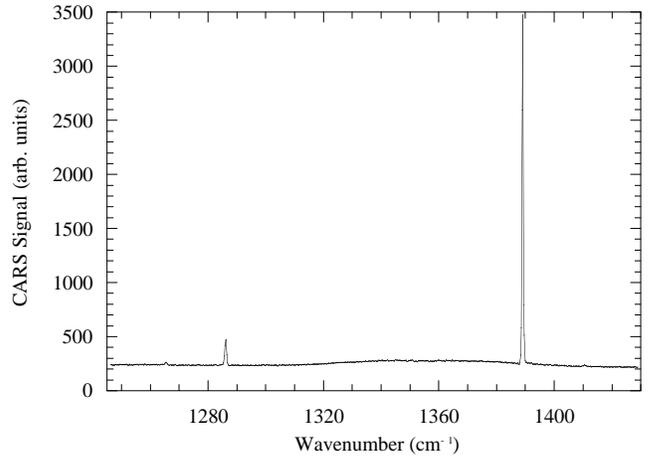}
\caption{Spectrum of carbon dioxide with discharge on.
		\label{fig:CO2.ldcdo}}
\end{figure}

As can be seen in Fig.~\ref{fig:CO2.ldcnd} and \ref{fig:CO2.ldcdo},
both peaks of the Fermi dyad drop when the discharge is turned on.
Surprisingly, the 1285~\cm\ line drops much more than the 1388~\cm\
line. This was true over broad ranges of both pressures and discharge
currents.

\subsection{Drop of Both Fermi Dyad Peaks \label{subsec:CO2both}}

The drop of {\em both} peaks when the discharge is turned on is
expected because the gas in the discharge is hotter and less dense
than the gas with no discharge. It is also vibrationally heated,
leading to the depletion of the ground state. In their report of
spontaneous Raman investigations of the pure rotational lines in
\carb\ Barrett~\et~\cite{Barrett:discharge} stated the intensity
of the lines with their discharge on (run at 10~mA and a pressure of
40~Torr) was about $\frac{1}{5}$ the intensity with the discharge off.
They calculated the temperature of their discharge (from the strongest
pure rotational line) to be 730~K.

In this experiment, such a measurement is not possible. The
temperature can be estimated, however, by calculating the relative
signal strength of the 1285~\cm\ and 1388~\cm\ peaks to the
1265~\cm\ and 1410~\cm\ hotband peaks as a function of temperature
and then matching the ratio with what is observed.

The population of a specific state is given by,
\begin{equation}
N_{i} = N \; \frac{g_{i} \, e^{-\frac{E_{i}}{kT}}}{Q_{v}(T)} \; ,
\label{eqn:pop}
\end{equation}
where, $N$ is the total number of atoms,
	$N_{i}$, $g_{i}$, and $E_{i}$ are the population,
	degeneracy factor, and the energy of the $i^{\textrm{th}}$ state
	and $Q_{v}(T)$ is the vibrational partition function.

The CARS signal is proportional to the square of the difference
between two states, $(\Delta N_{12})^2$, so the ratio of the intensity
of the ground state peaks to that of the hotband peaks is given by,
\begin{equation}
\left ( \frac{N_0 - N_2}{N_1 - N_3} \right ) ^2 =
	\left ( \frac{
		g_{0} - g_{2} \, e^{-\frac{E_{2}}{kT}}
	}{
		g_{1} \, e^{-\frac{E_{1}}{kT}} -
		g_{3} \, e^{-\frac{E_{3}}{kT}}
	} \right ) ^2 \; .
\label{eqn:temp}
\end{equation}

\begin{figure}[hbt]  
\includegraphics[width=3.3in,height=!]{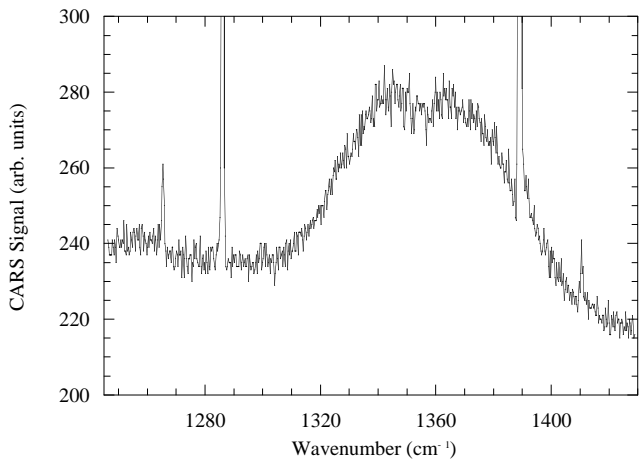}
\caption{Baseline of spectrum of Fig.~\ref{fig:CO2.ldcdo}
		showing hotbands.
\label{fig:CO2.ldcdob}}
\end{figure}

Fig.~\ref{fig:CO2.ldcdob} shows the detail of the baseline of
the discharge-on spectrum. The large hump in the middle of the
spectrum is background noise which remains when the cell is
evacuated and which may be a Raman signal from the optics.

The ratio of the summed areas of the 1285~\cm\ and 1388~\cm\
ground state peaks (7393) to the summed areas of the 1265~\cm\ and
1410~\cm\ hotband peaks (172) is 42.983. (Note that the baseline
was manually placed for each peak to compensate for the dispersive
parts of the lineshapes.) Solving Eqn.~\ref{eqn:temp} iteratively
for $T$ gives a discharge-on temperature of 373~K (quite a bit
lower than Barrett's 730~K). Note that this analysis assumes that
the mechanisms exciting a molecule from the ground state to one of
the hotbands are in thermal equilibrium with the mechanisms
de-exciting from the hotband to the ground state. As will be shown
in \S\ref{subsec:CO2vis}, this is an invalid assumption.

Nevertheless, this increased temperature would be expected to
reduce the density and cause a drop of signal strength by a factor
of 1.60 but there is an additional contribution from depletion of
the ground state by vibrational heating. To calculate this
contribution, it is necessary to use Eqns.~\ref{eqn:pop} and
\ref{eqn:temp} to calculate the exact ground state and excited state
populations for the discharge-off and discharge-on temperatures of
295~K and 373~K. Since the exponentials in $Q_{v}$ rapidly become
very small only a limited number of terms are needed. This
analysis uses the states shown in Fig.~\ref{fig:CO2el}, all of
which are below 3000~\cm. The calculation shows the ground state
population decreasing from 92\% of the molecules with the
discharge off to 67\% of the molecules with the discharge on and
25\% of the molecules being shifted from the ground state to
higher states. Including the density factor with the vibrational
heating, the 1285~\cm\ and 1388~\cm\ peaks should drop by a factor
of 1.9.

Since the 1285~\cm\ and 1388~\cm\ peaks decrease by more than a
factor of 19, much more than Barrett's factor of 5, there is an
additional source of decrease in $(\Delta N_{12})^2$.

One of the possibilities is that the ground state is being
depopulated because the \carb\ is being converted to other
molecules. Barrett, for example, considered that the drop in
signal might be due to the \carb\ molecules being converted to
CO$_2\/^+$. He estimated the concentration of these ions in the
discharge to be about $3 \times 10^{11}$~ions/cm$^3$, an
insignificant concentration compared to his gas density of
$10^{18}$/cm$^3$. Mass spectrometric measurements of discharges
have found other ions, such as O$_2\/^+$, but the total
concentrations were low
\cite{Austin:positive,%
Saporoschenko:drift,Johnson:instability,Kumar:CO2minus,%
Rees:transport}. This would also rule out large concentrations of
dissociation products.


\subsection{Greater Drop of 1285~\cm\ Peak \label{subsec:CO21285}}

In considering the greater drop of the 1285~\cm\ peak (10002)
than the 1388~\cm\ peak (10001), it must kept in mind that the
magnitude of the CARS signal depends upon the square of the
difference in population of the upper and lower states. To have
one peak drop more than the other, then, the ratio of the
differences in populations must be affected. This suggests
four possibilities:
\begin{itemize}		
\item{The transition rate is altered or the degeneracy of the
		harmonic states broken,}
\item{Raman saturation of upper levels,}
\item{The discharge saturates the final state of the
		1285~\cm\ transition, or,}
\item{The discharge changes the relaxation rates from
		the unperturbed states.}
\end{itemize}

Since both 10001 and 10002 begin in the ground state, it is
possible that the transition rate to 10002 is altered or the
near-degeneracy of the states $10^00$ and $02^00$ is changed.
Such, however, would require a shift of the wavenumber of the
peaks and none was observed (within a resolution of 0.2~\cm) when
the discharge was turned on. No change in the ratio of intensities
of 10001 and 10002 was also observed when the electric field was
applied without a discharge.

If the molecule is illuminated by strong enough laser beams, the
transition could be saturated with equal populations in the upper
and lower states. To test this, the YAG laser intensity was varied
by an order of magnitude. No significant change in the ratio of
intensities of 10001 and 10002 was observed.

The third possibility is not as easily addressed as the first two.
If the discharge saturates the population of 10002, its intensity
will be decreased relative to 10001. This would require an
electron excitation rate of $\Ket{\psi_l}$ in Eqn.~\ref{eqn:psiul}
greater than that of $\Ket{\psi_u}$. Zhu~\et~\cite{Zhu:quant.int},
however, observed exactly the opposite.

Their experimental setup was completely different than the one
used in this work and they worked at a much lower pressure
(27.5~milliTorr).

They produced translationally hot electrons, e$^{-}$*, by
multiphoton ionization of iodine with an ArF laser at 193~nm:
\[
\textrm{I}_2 + n\hbar\omega(193\textrm{nm}) \rightarrow
	\textrm{I}_2\/^+ + \textrm{e}^{-}\textrm{*} \; .
\]
These translationally hot electrons then collided with \carb\
molecules causing rotational and vibrational excitation of the
10001 and 10002 states. By tuning their diode laser, operating at
($\lambda \approx 4.3 \mu$m), they could probe specific rotational
lines of the \carb\ molecules by absorption in the strongly
allowed $\nu_3$ (anti-symmetric stretch) infrared band using the
transitions $10^00 \rightarrow 10^01$ and $02^00 \rightarrow 02^01$.
They concluded that the upper state, $\Ket{\psi_u}$ is significantly
more populated by electron scattering than the lower state,
$\Ket{\psi_l}$, by a factor of 9.1 to 12.7, depending on the
particular J state examined.

Looking at the collisional excitation probability from the ground
state to the upper excited state, P$_{gu}$, and to the lower
excited state P$_{gl}$,
\begin{gather*}
\textrm{P}_{\textrm{gu}} \propto
	\left|\, \sin\theta \Bra{10^00} V \Ket{00^00}
		+ \cos{\theta} \Bra{02^00} V \Ket{00^00} \, \right|^2 \\
\textrm{P}_{\textrm{gl}} \propto
	\left|\, \cos\theta \Bra{10^00} V \Ket{00^00}
		- \sin{\theta} \Bra{02^00} V \Ket{00^00} \, \right|^2
\end{gather*}
where $V$ defines the interaction potential. If one assumes the
electron excitations $\Bra{10^00} V \Ket{00^00}$ and $\Bra{02^00}
V \Ket{00^00}$ are nearly equal, Zhu's results follow from these
probabilities. The minus sign in the equation for $\ket{\psi_l}$
then causes a ``quantum interference'' in its electron excitation
rate, causing the upper state of $\Ket{\psi_l}$ to have a lower
population than the upper state of $\Ket{\psi_u}$. This is in
contrast to the present observations where $\Ket{\psi_l}$'s weaker
signal implies a higher population in its upper state.
 Since one would expect that a change in population
that affects the infrared signal in one way would not affect the
CARS signal in another, there must be a difference in the
experiments to cause different results.

The difference is that Zhu~\et\ used ``hot'' electrons, a
significant fraction of which had energies around 3~eV, while our
experiment produces mean electron energies on the order of 0.1~eV.
Their higher energy electrons opened up an excitation channel that
was unavailable to the lower energy electrons of this experiment.

\begin{figure}[htb]  
\includegraphics[width=3.3in,height=!]{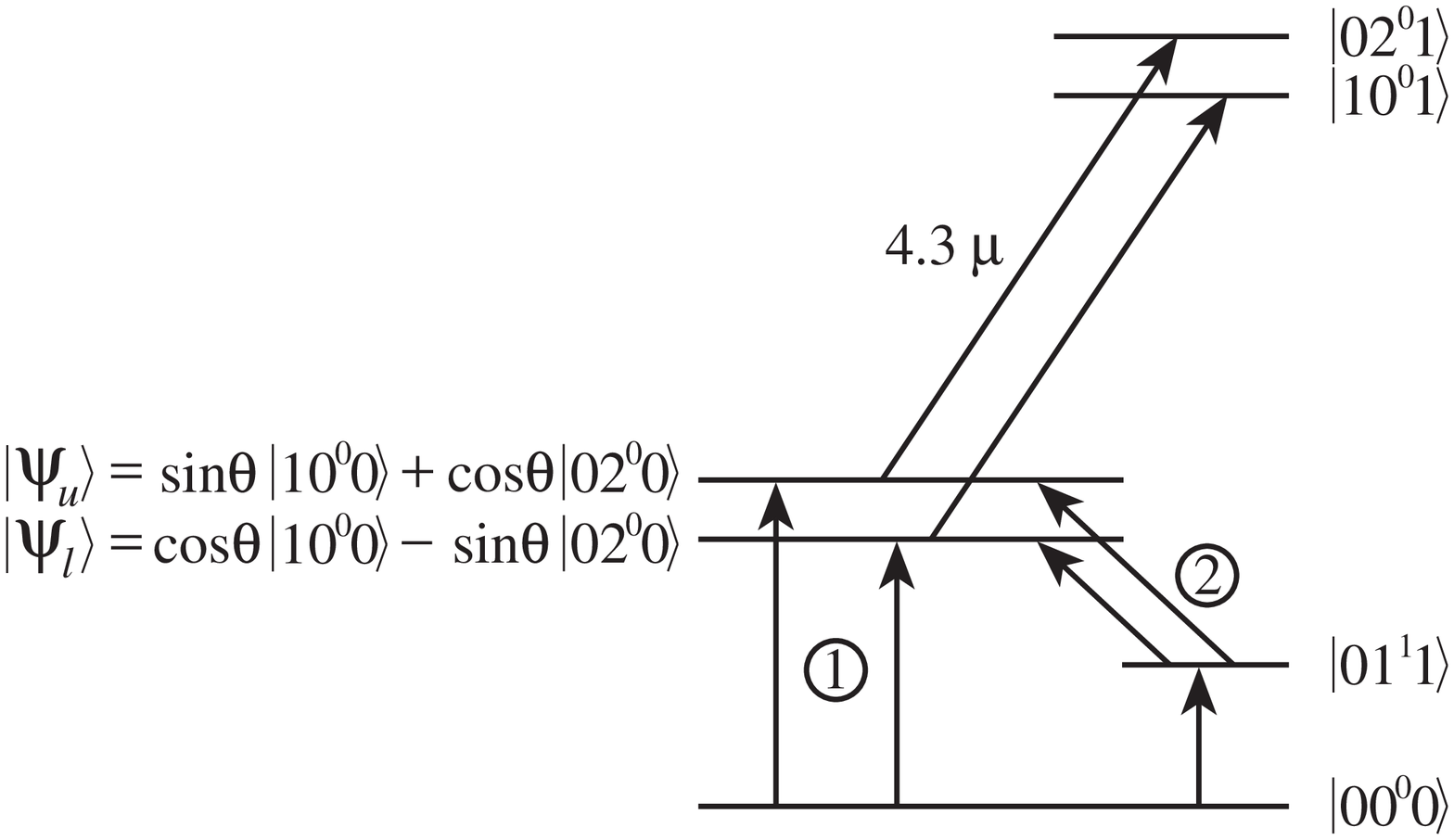}
\caption{\carb\ excitation paths. \label{fig:CO2.excitation}}
\end{figure}

Counting as separate the unperturbed states making up
$\Ket{\psi_l}$ and $\Ket{\psi_u}$, there are four paths of
excitation from the ground state to $\Ket{\psi_l}$ and
$\Ket{\psi_u}$. As shown in Fig.~\ref{fig:CO2.excitation}, there
are two direct paths marked with the circled ``1'',
\begin{eqnarray*}
00^00 & \rightarrow & 10^00	\\
00^00 & \rightarrow & 02^00	\; ,
\end{eqnarray*}
and two indirect paths marked with the circled ``2'',
\begin{eqnarray*}
00^00 & \rightarrow \quad \!\! 01^10 & \rightarrow \quad \!\! 10^00  \\
00^00 & \rightarrow \quad \!\! 01^10 & \rightarrow \quad \!\! 02^00  \; .
\end{eqnarray*}

The single quantum, direct excitation of $10^00$ from $00^00$ occurs
most readily in the discharge. Mazevet~\et~\cite{Mazevet:virtual}
showed that the vibrationally inelastic excitation cross-section for
e-\carb\ scattering peaks around 0.2~eV, quite close to the average
electron energy in the discharge of 0.1~eV mentioned in the previous
section.

The double quantum, direct excitation of $02^00$ from $00^00$ is
through the intermediate state of CO$_2\/^-$. CO$_2\/^-$, which
can have a lifetime as long as 90~$\mu s$ in the gas phase
\cite{Compton:collision}, has a bond angle of 134.9\textdegree\
\cite{Rossi:structure}), so the bend vibration is strongly excited
through the decay of CO$_2\/^-$. The potential energy curves of
CO$_2\/^-$ are, however, roughly 3~eV higher than those of \carb\
\cite{Spence:cross,Claydon:scatt}, so this excitation channel was
open to Zhu~\et\ but closed in the present experiment.

Both indirect paths go through the singly excited bend state
$01^10$ as an intermediate. Its excitation energy of 667~\cm\
(0.083~eV) is matched to the average electron energy in the
discharge. The second half of the first indirect path, $00^00
\rightarrow 01^10 \rightarrow 10^00$, is a two quantum transition
--- it requires a change of v$_1$ from zero to one and a
simultaneous change of v$_2$ from one to zero --- whose
probability is therefore low. The second half of the second
indirect path, $00^00 \rightarrow 01^10 \rightarrow 02^00$, is a
single quantum transition and requires nearly the same energy as
the first half of the indirect paths. It thus has a high
probability and is the route for excitation of $02^00$ in this
experiment.

The difference between the results seen by Zhu~\et's experiment
and the present one is that Zhu~\et\ had direct excitations from
the ground state, $00^00$, to the final state, $\Ket{\psi_l}$ or
$\Ket{\psi_u}$, through two coherent channels, one exciting
$10^00$ and the other exciting $02^00$. Because these two channels
can occur coherently during a single electron excitation process,
they saw interference between them and thus a lower excitation
probability for the production of $\Ket{\psi_l}$. The present
experiment, using ``cool'' electrons, in contrast, has direct
excitation $00^00 \rightarrow 10^00$, but indirect, two-step
excitation $00^00 \rightarrow 01^10 \rightarrow 02^00$. Since
these two processes are incoherent there is no interference and
the excitation probabilities of $\Ket{\psi_l}$ and $\Ket{\psi_u}$
are similar, leading to the conclusion that the discharge could
not have preferentially saturated one of the observed states.

The fourth and remaining possibility for explaining this result is
that the discharge changes the relaxation rates from the
unperturbed states. Here the minus sign of $\Ket{\psi_l}$ in
Eqn.~\ref{eqn:psiul} does explain what was seen. If turning on the
discharge causes the transition rates for the unperturbed states,
$10^00$ and $02^00$, to become more equal, the quantum
interference engendered by the minus sign in Eqn.~\ref{eqn:psiul}
will cause $\Ket{\psi_l}$ to have a much smaller relaxation rate
to the lower levels, $01^10$ and $00^00$, than $\Ket{\psi_u}$. (It
can readily be seen that the relaxation rates can be different
between discharge on and discharge off because when the discharge
is off, relaxation occurs by spontaneous emission and inelastic
collisions with molecules and the walls (collision induced
emission). When the discharge is turned on, relaxation
additionally occurs by inelastic and superelastic collisions with
free electrons and ion and electron charge ($\vec{E}$~field)
induced emission.)

The significant drop of the 1285~\cm\ peak can thus be explained
by the discharge causing changes in the relaxation rates from the
unperturbed states and saturating the state $\ket{\psi_l}$. Note
that the majority of observed drop of the 1388~\cm\ line could
also be due to the state $\ket{\psi_u}$ being partially saturated
by the discharge.

\begin{figure}[th]  
\includegraphics[width=3.3in,height=!]{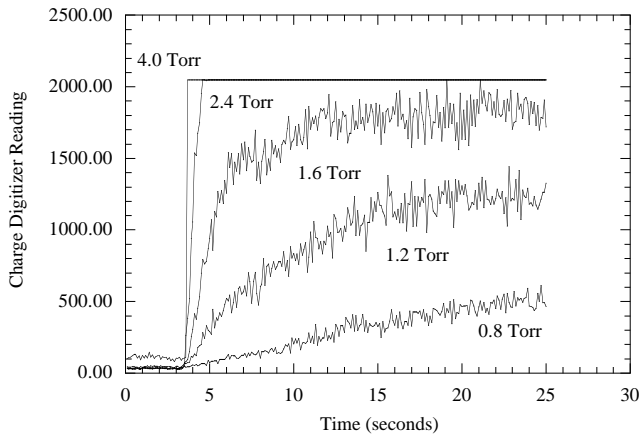}
\caption{Rise of 1285~\cm\ signal of \carb\ with turn-off of discharge.
\label{fig:CO2.1285.rise}}
\end{figure}

\begin{figure}[thb]  
\includegraphics[width=3.3in,height=!]{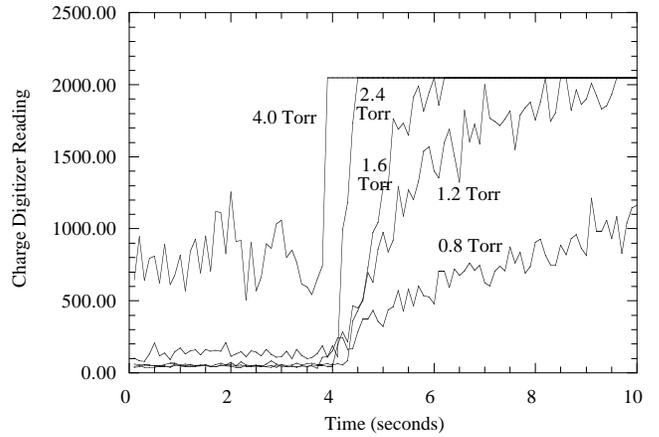}
\caption{Rise of 1388~\cm\ signal of \carb\ with turn-off of discharge.
\label{fig:CO2.1388.rise}}
\end{figure}

To test this hypothesis, the change of 1285~\cm\ and 1388~\cm\ CARS
signals as a function of time after the discharge was turned off was
measured for pressures between 0.8 to 4~Torr. These data are graphed
in Fig.~\ref{fig:CO2.1285.rise} and \ref{fig:CO2.1388.rise}.

For a pressure of 1.6~Torr, for example, the 1388~\cm\ signal
increased as an exponential with a time constant $\tau\approx
2.0\pm0.3$~s. The increase in signal is due to both the decrease
of population in $\Ket{\psi_u}$ and the repopulation of the ground
state from other levels by collisions. This vibrational relaxation
time for repopulating the ground state decreases with increasing
pressure as expected.

\begin{figure}[htb]  
\includegraphics[width=3.3in,height=!]{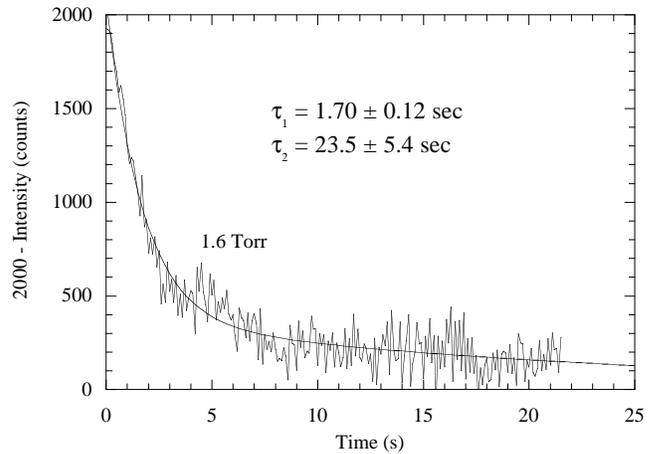}
\caption{Rise of 1285~\cm\ signal of \carb\ at 1.6~Torr with double
		exponential fit.
\label{fig:CO2.rise.fit}}
\end{figure}

Under the same conditions, the growth of the 1285~\cm\ signal due to
$\Ket{\psi_l}$ exhibits a double-exponential behavior whose shorter
time constant is consistent with the repopulation of the ground state.
These data are re-plotted in Fig.~\ref{fig:CO2.rise.fit} with the
curve fit.
The results of this analysis are consistent with the model of quantum
interference altering the vibrational relaxation rate from $\Ket{\psi_l}$
compared with that from $\Ket{\psi_u}$.

Others have measured relaxation rates from \carb\ states but their
results are not applicable to this experiment.
Roche~\et~\cite{Roche:relax}, for example, measured relaxation
  of the $\nu_1/2\nu_2$ Fermi dyad with Raman-infrared double
  resonance using a \carb\ laser to continuously
  monitor populations via the 9.4 or 10.4~$\mu$m \carb\ laser
  transitions much in the same way as Zhu~\et used their diode
  laser. Instead of exciting \carb\ molecules with hot electrons
  or H(D) atoms, they selectively excited the molecules with a
  doubled-YAG/dye laser setup similar to the one used in the
  present work ---
  though theirs achieved narrow-band dye operation with an argon laser
  pumped CW dye oscillator. They measured the vibrational relaxation
  rate versus pressure from the 1388~\cm\ peak. Since the 1388~\cm\
  peak is very narrow the rotational lines blend at low pressures
  (0.006 amagat, 5~Torr). Because of this, they excited all of the
  rotational lines, which lead to achieving rotational equilibrium
  more rapidly and making the vibrational relaxation more efficient.
  They were able to resolve the 1285~\cm\ Q-branch and used that
  to measure the rotational relaxation rate versus pressure from
  that state.  Unfortunately, they do not give the vibrational
  relaxation rate from the 1285~\cm\ peak and do not give any
  results for \carb\ in a discharge.
Dang~\et~\cite{Dang:dynamics} measured the relaxation times of \carb\
  vibrational levels in a discharge. They disturbed the discharge-on
  equilibrium with a \carb\ laser pulse and then watched as that
  equilibrium was re-established with the discharge remaining on.
In this experiment, in contrast, the discharge-on relaxation rates
  are compared with the discharge-off relaxation rates and a change
  in the rates when the discharge is turned on is what causes the
  1285~\cm\ peak to drop more than the 1388~\cm\ peak.


\subsection{Visibility of Saturation of $\Ket{\psi_l}$ in Spectra%
		\label{subsec:CO2vis}}

If, as posited in the previous section, the 1285~\cm\ signal drops
more than the 1388~\cm\ signal because its lower relaxation rate
causes a saturation of the upper level of the transition, that
saturated upper level should then be the lower level of hotband
transitions that appear when the discharge is turned on. The
possible transitions out of both the 1285~\cm\ and 1388~\cm\
states with energies similar to those of the present investigation
are shown in Table~\ref{table:hotbands}.
\begingroup
\begin{table}[hbt]
\caption{Hotband transitions out of 1285~\cm\ and 1388~\cm\ states.
		\label{table:hotbands}}
\begin{ruledtabular}
\begin{tabular}{cccc}
      &                         &\multicolumn{2}{c}{ }             \\[-10pt]
      &                         &\multicolumn{2}{c}{Lower State}   \\
      &                         & 10001         & 10002            \\
\multicolumn{2}{c}{Upper State} & (1388.18~\cm) & (1285.41~\cm)    \\
      &                         &               &                  \\[-10pt]
\cline{1-4}
      &                         &               &                  \\[-10pt]
20003 & (2548.37~\cm)           & 1160.18       & 1262.96          \\
12202 & (2585.02~\cm)           & 1196.84       & 1299.61          \\
20002 & (2671.14~\cm)           & 1282.96       & 1385.73          \\
12201 & (2760.72~\cm)           & 1372.54       & 1475.32          \\
20001 & (2797.14~\cm)           & 1408.95       & 1511.73          \\[3pt]
\end{tabular}
\end{ruledtabular}
\end{table}
\endgroup
Of the ten candidates, six are within the 1245~\cm\ to 1430~\cm\
range examined experimentally. Of these six, the three from
1285~\cm\ should be stronger than the three from 1388~\cm. In
addition, two of the transitions ($12201\leftarrow10001$ at
1372.54~\cm\ and $12202\leftarrow$10002 at 1299.61~\cm) involve
the change of only the number of quanta of the bend vibration,
$\nu_2$. As mentioned in \S\ref{subsec:CO2anharm}, the $\nu_2$
vibration is Raman inactive and so, absent any concomitant change
in the symmetric stretch ($\nu_1$) quantum number or Fermi mixing,
should not be visible at all.

The baselines of many discharge-on spectra were examined for
evidence of lines at these wavenumbers. Of the six candidates,
1299.61~\cm, 1385.73~\cm, and 1408.95~\cm\ suggest the
possibilities of lines.

\begin{figure}[thb]  
\includegraphics[width=3.3in,height=!]{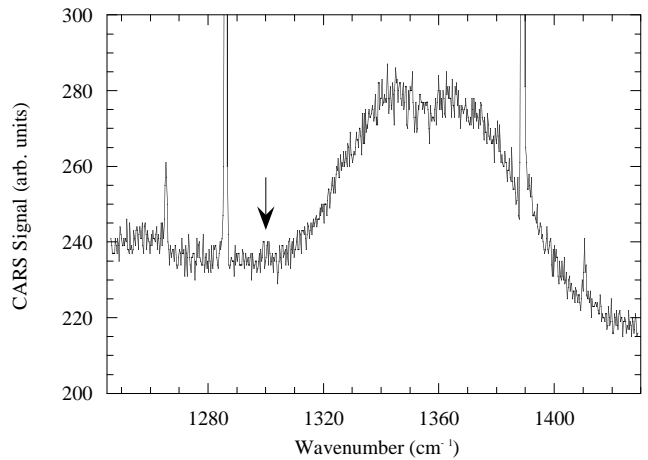}
\caption{Baseline of \carb\ spectrum showing 1299.61~\cm\ line
	when discharge is turned on. \label{fig:CO2hothotbands.1299.61}}
\end{figure}

The spectrum examined previously for computation of the
temperature of the discharge (Fig.~\ref{fig:CO2.ldcdob}) is
reproduced in Fig.~\ref{fig:CO2hothotbands.1299.61}
with an arrow pointing at the possible line at 1299.61~\cm,
heavily influenced by noise. Its weakness is not surprising in
view of the Raman inactivity of the transition.

\begin{figure}[bht]  
\includegraphics[width=3.3in,height=!]{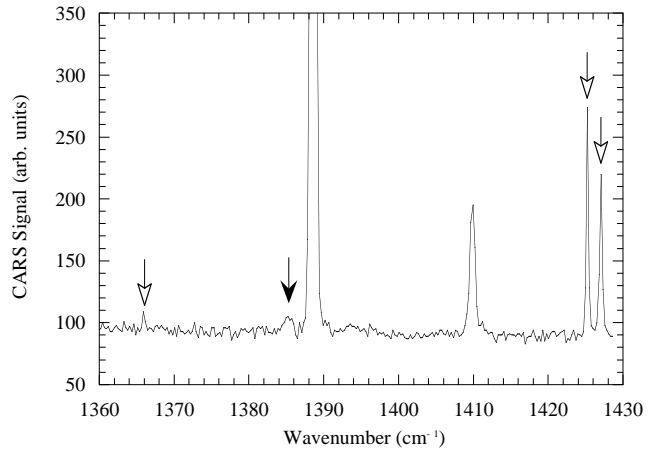}
\caption{Baseline of \carb\ spectrum showing 1385.73~\cm\ line
	(solid arrow) and additional lines at 1366, 1424, and 1426~\cm\
	when the discharge is turned on (hollow arrows).
\label{fig:CO2additional}}
\end{figure}

Fig.~\ref{fig:CO2additional} shows the region of the 1388~\cm\
Q-branch with the solid arrow pointing at the possible line at
1385.73~\cm. The weakness of this line is surprising given the
degree of saturation expected for the upper state of the 1285~\cm\
Q-branch, $\ket{\psi_l}$, though it could be explained by a small
Raman cross-section for the transition.
It is also possible that the that the cross-products between a
line and its neighbors and between a line and the background
inherent in the $|\chi^{\,(3)}|^2$ probed by CARS destructively
interfere and reduce the strength of the line.

\begin{figure}[hbt]  
\includegraphics[width=3.3in,height=!]{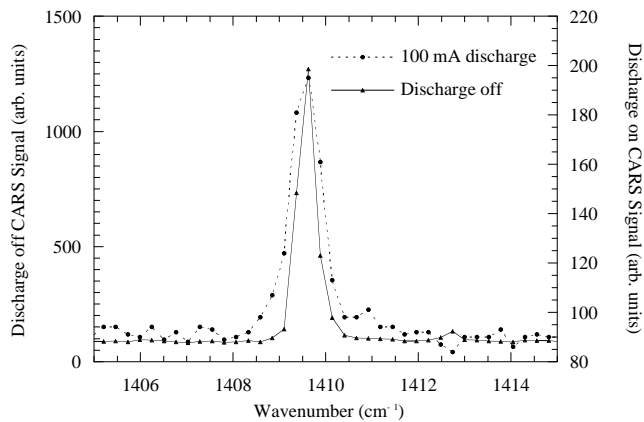}
\caption{Baseline of \carb\ spectra showing comparison of 1409.476~\cm\
		hotband with discharge off (solid line) and on (dotted line).
\label{fig:CO2hothotbands.1409}}
\end{figure}

Fig.~\ref{fig:CO2hothotbands.1409}
shows a comparison of the no discharge and 100~mA discharge
spectra of $11101\leftarrow01101$ hotband at 1409.476~\cm. Note
that the discharge on peak is broader and is less symmetric than
the discharge off peak, being shaded to the lower energy side,
where the $20001\leftarrow10001$ hotband would be. In contrast to
the amplitude of the 1385~\cm\ line just discussed, here there is
no question that the cross-products in the $|\chi^{\,(3)}|^2$
change the shape and placement of the lines. This effect could be
avoided by using RIKES instead of CARS. Since this line arises
from the upper state of the 1388~\cm\ Q-branch, $\ket{\psi_u}$,
its weakness is expected.

Interestingly, however, it was noticed that when the discharge was
on, there were repeatable, clearly resolvable lines at 1366, 1424,
and 1426~\cm. These lines are marked in Fig.~\ref{fig:CO2additional}
with hollow arrows.
From Rothman's data \cite{Rothman:energylevels},
the 1366~\cm\ line is due to the transition $10011\leftarrow00011$,
the 1424~\cm\ line to $21101\leftarrow11101$, and
the 1426~\cm\ line to $12201\leftarrow02201$,
where 02201 is the two
quanta bend vibrational state with two quanta of vibrational angular
momentum that is not involved in Fermi resonance with 10001.
Though hotband transitions directly from the saturated 10001 and 10002
states are not observed, the strong hotband from 02201 only 50~\cm\
away suggests large populations of all three levels. This validates
the saturation model.

Note that in Fig.~\ref{fig:CO2additional}, the amplitude of the
line at 1424~\cm\ (which originates from the 02201 state at
1335.15~\cm) is greater than that of the line at 1410~\cm\ (which
originates from the 01101 state at 667.3799~\cm), whereas a
Boltzman distribution at 373~K would require the opposite.
This indicates that the gas temperature calculated in
\S\ref{subsec:CO2both} from the ratio of the populations is
incorrect, that the vibrational energy of the molecules is not in
thermal equilibrium with the temperature of the gas but is more on
the order of the temperature of the electrons, which for energies
of 0.1~--~0.2~eV is roughly 1000~--~2000~K. This is not surprising
because the molecules are excited to upper levels by electron
impact while they are de-excited back to the ground state by
collisions with other molecules. In fact, with separate processes,
one would be surprised if the excitation and de-excitation rates
were equal and thermal equilibrium were achieved.

\begin{figure}[b]  
\includegraphics[width=3.3in,height=!]{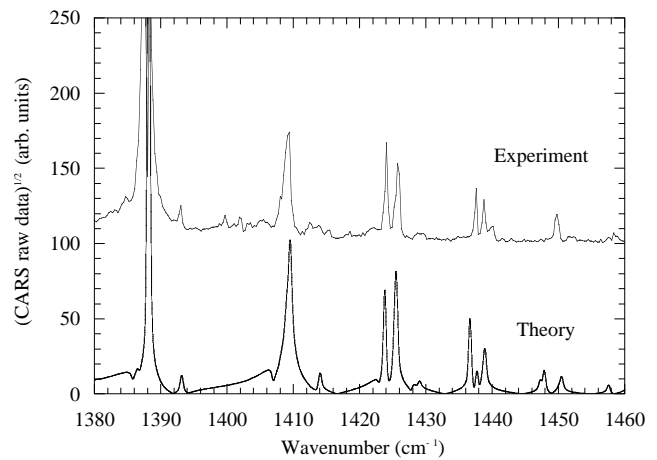}
\caption{High pressure \carb\ spectra showing higher-order hotbands. Note
		that the 1388~cm$^{-1}$ peak in the experimental (upper)
		trace is saturated; the strong signals at 40~Torr overload
		the electronics. \label{fig:CO2highpressure}}
\end{figure}

In contrast to the very selective population of upper levels by
electron impact evident in low pressure spectra, spectra taken at
higher pressures show a much more thermal distribution. In
Fig.~\ref{fig:CO2highpressure},
the intensities of the excited state lines show a monotonic decrease
in strength, exactly what one would expect from a Boltzman distribution
and similar to what has been shown previously for work in flames
\cite{Blint:flames} and in furnaces \cite{farrow:glass}; the quantum
interference in the 1285 and 1388~\cm\ lines is still apparent.

Quantum interference is present in other transitions of \carb\ as
well. Table~\ref{table:hotbands} shows that a line at 1511~\cm\
due to the transition $20001 \leftarrow 10002$ should be observed.
Since this transition is due to an increase of the Raman-active
symmetric stretch quantum number, one would expect it to be quite
visible. Experimentally, however, this line is not seen, even in
spectra at 40~Torr. In this case, the level 20001 is composed of
the states $20^00 + 12^00 + 04^00$ and the level 10002 is composed
of the states $10^00 - 02^00$. Here excitation by the path $12^00
\leftarrow 02^00$ (with the minus sign) interferes with excitation
by the path $20^00 \leftarrow 10^00$ and the total excitation
probability vanishes. It is also interesting to note that the
1409~\cm\ hotband, perhaps seen only as a broadening in
Fig.~\ref{fig:CO2hothotbands.1409}, is more distinct as a separate
peak in Fig.~\ref{fig:CO2highpressure}.


\bibliography{bibliography}

\end{document}